\documentclass[12pt]{article}
\usepackage{epsfig}

\input{tcilatex}

\begin{document}

\title{Description of the Ground and Octupole Bands \\
in the Symplectic Extension of the Interacting Vector Boson Model}
\author{H. Ganev, V. P. Garistov, A. I. Georgieva \\
\textit{Institute of Nuclear Research and Nuclear Energy,}\\
\textit{\ Bulgarian Academy of Sciences, Sofia, Bulgaria}}
\maketitle

\begin{abstract}
In the framework of the symplectic \ extension \ of the \ Interacting Vector
Boson Model (IVBM) a good description of the first \ excited positive \ and
negative parity bands of the nuclei in the rare earth and the actinide
region is achieved. \ The bands investigated in the model are extended to
very high angular momenta as \ a result of their consideration as "yrast"
bands \ with respect to the symplectic \ classification \ of \ the \ basis
states. The \ analysis of the eigenvalues of the \ model Hamiltonian \
reveals the presence \ of an interaction between these \ bands. Due to this
iteraction the $\Delta L=1$ staggering effect between the energies of \ the\
states of two bands is also reproduced including \ the "beat" patterns.
\end{abstract}

\section{Introduction}

The existence of nuclei with stable deformed shapes was realized early in
the history of nuclear physics. The observation of large quadrupole moments
led to the suggestion that some nuclei might have spheroidal shapes, which
was confirmed by the observation of rotational band structures and
measurements of their properties. For most deformed nuclei, a description as
an axial- and reflection-symmetric spheroid is adequate to reproduce the
band's spectroscopy. Because such a shape is symmetric under space
inversion, all members of the rotational band have the same parity. However,
with the first observation of negative parity states near the ground state,
the possibility arose that some nuclei might have an asymmetric shape under
reflection.

On the other hand, whenever symmetry breaking appears new behavior of the
many-body system is expected. Reflection symmetry breaking is associated
with a static octupole deformation which is expected to determine new
collective features for the nuclear system.

Extensive investigations into the structure of nuclei with low-lying
negative parity states has led to the conclusion that, while reflection
asymmetric shapes can play a role in the band structure, they are not as
stable as the familiar quadrupole deformations. The rotational spectra of
some even-even nuclei in the rare earth and light actinide region exhibit,
next to the ground band, a negative parity band which consists of the states
with $I^{\pi }=1^{-},3^{-},5^{-},...$ . These two bands are displaced from
each other, which means that fluctuations back to space symmetric shapes
must also be significant. Experimentally \ the presence of "octupole" bands
for some isotopes from the light actinide and rare earth region \cite{revexp}
is firmly established.

There is a large variety of models that try to describe this behaviour of
the low-ling states of deformed nuclei. Particularly successful are
algebraic models based on symmetry principles. The introduction of an
additional octupole degrees of freedom is a common feature of \ most of
those models .

The prescription for describing negative parity states by the addition of an 
$f$ \ boson to the usual $s$ and $\ d$ of the IBM was first mentioned by
Iachello and Arima \cite{IBM}. It was suggested \cite{Iach2} that the
inclusion of a $p$ boson to the $s,d$ and $f$ bosons may play an important
role in the \ description of these \ collective \ states.

The coherent state method (CSM) was applied by Alonso et al. to the \textit{%
spdf} $SU(3)$ Hamiltonian with quadrupole and octupole interaction \cite%
{alonso}. Recently\ A. A. Raduta and D. Ionescu \cite{raduta} have used a
generalization of the CSM . They suggested that both ground and octupole
bands may be considered as being projected from a single deformed intrinsic
state that exhibits both quadrupole and octupole deformations.

Another collective \ model based on point symmetry group considerations \cite%
{NS1}\ has also been used very successfully for the description of the
energy levels of the ground and octupole bands and reproduces odd-even
staggering between these levels \cite{NS2}. In this model the octupole field
is parametrized by irreducible representations of the octachedron point
symmetry group.

The introduction of an octupole degrees of freedom in the presence of
comparatively large number of free parameters in all of these models allows
for the reproduction of the experimental data of the energies of the
negative parity states, at least in the low spin region.

In the beginning of the 1980's a phenomenological algebraic model called the
Interacting Vector Boson Model (IVBM) was introduced \cite{IVBMb}. This
model is a generalization of \ the phenomenological broken $SU(3)$ symmetry
model \cite{RuRa}, which provided a good description of the low-lying ground
\ and \ $\gamma $ \ bands \cite{NS3}\ of well deformed even-even nuclei. Its
advantages were incorporated into \ the rotational limit of the IVBM \cite%
{IVBMrl}, with an a good description of all the positive parity bands of
nuclei in the rare earth and the actinide region. Moreover, the $U(6)$
extension of the model contains such sequences of \ $SU(3)$ multiplets, some
of which proves to be convenient for the description of the low-lying
negative parity bands \cite{oepb} .

With the recent advance of the experimental technique the investigated bands
were extended to very high angular momenta \cite{revexp}. This motivated a
new approach within the framework of the model aimed at a description of the
first positive and negative bands, up to very high spins. In this new
application, we make use of the symplectic extension of the IVBM \cite{ggg}.
This allows these bands to be considered as yrast bands in the sense, that
we take into account the states with a given $L,$ which minimize the energy
values with respect to $N$. $N$ is the eigenvalue of the total number of
bosons that build the basis states of the model. Its eigenvalue changes \ as 
$\Delta N=2$ in the infinite spaces of the boson representation of $Sp(12,R)$%
. When considering the dynamical symmetry of the symplectic extension of the
model through the maximal compact subgroup $U(6)\supset Sp(12,R)$, we obtain
the exactly solvable rotational limit with a Hamiltonian, diagonal in a
basis defined by the irreducible representations of the corresponding chain
of subgroups. The measured energies of the ground and octupole bands in
even-even nuclei from the rare earth and actinide regions are reproduced in
the model with rather good accuracy. The analysis of the obtained results
shows that this is due to the appearance of a vibrational type term \ that
influences the yrast energies. This term also plays the role of an
interaction between the two considered bands, and is \ the reason \ for the
correct reproduction of the odd-even \ staggering \ of \ their energies.

\section{Algebraic Basis of the IVBM.}

We start with a brief review of the model's assumptions and definitions. The
IVBM is based on the introduction of two kinds \ of vector bosons (called $p$
and $n$ bosons), that \textquotedblleft built up\textquotedblright\ the
collective excitations in the nuclear system. The creation operators of
these bosons are assumed to be $SO(3)$ vectors and they transform according
to two independent fundamental representations (1,0) of the group $SU(3)$ .
These bosons form a \textquotedblright pseudospin\textquotedblright\ doublet
of the $U(2)$ group and differ in their \textquotedblleft
pseudospin\textquotedblright\ projection $\alpha =\pm \frac{1}{2}.$ \ The
introduction of this additional degree of freedom leads to the extension of
the $SU(3)$ symmetry to $U(6)$. Then the operators

\begin{equation}
u_{m}^{+}(\alpha =\frac{1}{2})=p_{m}^{+},\text{ \ \ \ \ \ }u_{m}^{+}(\alpha
=-\frac{1}{2})=n_{m}^{+}  \label{bosons}
\end{equation}
transform according to the fundamental representation $[1]_{6}$ of the group 
$U(6)$. The annihilation operators are obtained by the conjugation $%
(u_{m}^{+}(\alpha ))^{\dagger }=u_{m}(\alpha )$ and transform according to
the conjugate $SU(3)$\ representations (0,1)$.$ The bilinear products of the
creation and annihilation operators of the two vector bosons generate the
noncompact symplectic group $Sp(12,R)$ \cite{IVBMb}: 
\[
F_{M}^{L}(\alpha ,\beta )=_{k,m}^{\sum }C_{1k1m}^{LM}u_{k}^{+}(\alpha
)u_{m}^{+}(\beta ), 
\]

\[
G_{M}^{L}(\alpha ,\beta )=_{k,m}^{\sum }C_{1k1m}^{LM}u_{k}(\alpha
)u_{m}(\beta ), 
\]

\begin{equation}
A_{M}^{L}(\alpha ,\beta )=_{k,m}^{\sum }C_{1k1m}^{LM}u_{k}^{+}(\alpha
)u_{m}(\beta ),  \label{generators}
\end{equation}
where $C_{1k1m}^{LM}$ are the usual Clebsh-Gordon coefficients and $L$ and $%
M $ define the transformational properties of (\ref{generators}) under
rotations.

We consider $Sp(12,R)$ to be the group of the dynamical symmetry of the
model \cite{IVBMb}. Hence the most general one and two-body Hamiltonian can
be expressed in terms of its generators . Using commutation relations
between the $F_{M}^{L}(\alpha ,\beta )$ and $G_{M}^{L}(\alpha ,\beta )$, the
full range of number of bosons preserving Hamiltonian can be expressed in
terms of operators $A_{M}^{L}(\alpha ,\beta )$: 
\begin{equation}
H=\sum_{\alpha ,\beta }h_{0}(\alpha ,\beta )A^{0}(\alpha ,\beta
)+\sum_{M,L}\sum_{\alpha \beta \gamma \delta }(-1)^{M}V^{L}(\alpha \beta
;\gamma \delta )A_{M}^{L}(\alpha ,\gamma )A_{-M}^{L}(\beta ,\delta ),
\label{general hamiltonian}
\end{equation}
where $h_{0}(\alpha ,\beta )$ and $V^{L}(\alpha \beta ;\gamma \delta )$ are
phenomenological constants.

Being a noncompact group, the representations of $Sp(12,R)$ are of infinite
dimension, which makes it impossible to diagonalize the most general
Hamiltonian. The operators $A_{M}^{L}(\alpha ,\beta )$ generate the maximal
compact subgroup of $Sp(12,R)$, namely the group $U(6)$: 
\[
Sp(12,R)\supset U(6) 
\]
So the even and odd unitary irreducible representations (UIR) of $Sp(12,R)$
split into an infinite but countable number of symmetric UIR of $U(6)$ of
the type $[N,0,0,0,0,0]=[N]_{6}$, where $N=0,2,4,...$ for the even set (see
Table 1) and $N=1,3,5,...$ for the odd set \cite{sp4nR}. These subspaces are
of finite dimension, which simplifies the problem of diagonalization.
Therefore the \textit{complete} spectrum of the system can be calculated
only trough the diagonalization of the Hamiltonian in the subspaces of 
\textit{all }the UIR of $U(6)$, belonging to a given UIR of $Sp(12,R)$.

The rotational limit \cite{IVBMrl} of the model is further defined by the
chain:

\begin{equation}
U(6)\supset SU(3)\otimes U(2)\supset SO(3)\otimes U(1)  \label{chain}
\end{equation}
\begin{equation}
\lbrack N]\ \ \ \ \ \ (\lambda ,\mu )\ \ \ \ \ (N,T)\ \ K\ \ \ \ \ L\ \ \ \
\ \ \ \ \ \ T_{0}  \label{qnum}
\end{equation}
where the labels below the subgroups are the quantum numbers (\ref{qnum})\
corresponding to their \ irreducible representations. Their values are
obtained by means of standard reduction rules and are given in \cite{IVBMrl}%
. In this limit the operators of the physical observables are the angular
momentum operator 
\[
L_{M}=-\sqrt{2}\sum_{\alpha }\ A_{M}^{1}(\alpha ,\alpha ) 
\]
\bigskip and the truncated (``Elliott'')\ quadrupole operator 
\[
Q_{M}=\sqrt{6}\sum_{\alpha }A_{M}^{2}(\alpha ,\alpha ), 
\]
which define the algebra of $SU(3)$.

The operators \ of \ the \ \textquotedblleft pseudospin\textquotedblright\
components and the number of bosons $N$ : 
\begin{eqnarray*}
T_{+1} &=&\sqrt{\frac{3}{2}}A^{0}(p,n);\ \ \text{\ \ \ \ \ \ \ \ \ \ \ \ \ \
\ \ \ \ \ \ \ \ \ \ \ \ \ \ }\ T_{-1}=-\sqrt{\frac{3}{2}}A^{0}(n,p); \\
T_{0} &=&-\sqrt{\frac{3}{2}}[A^{0}(p,p)-A^{0}(n,n)];\ \ \text{\ \ \ \ \ \ \ }%
\ N=-\sqrt{3}[A^{0}(p,p)+A^{0}(n,n)],
\end{eqnarray*}
define the algebra of $U(2)$.

Since the reduction from $U(6)$ to $SO(3)$ is carried out by the mutually
complementary groups $SU(3)$ and $U(2)$, their quantum numbers are related
in the following way: 
\begin{equation}
T=\frac{\lambda }{2},\text{ \ \ \ \ \ \ }N=2\mu +\lambda  \label{NTcon}
\end{equation}
Making use of the latter we can write the basis as 
\begin{equation}
\mid \lbrack N]_{6};(\lambda ,\mu );K,L,M;T_{0}\rangle =\mid
(N,T);K,L,M;T_{0}\rangle  \label{bast}
\end{equation}
The ground state of the system is: 
\begin{equation}
\mid 0\text{\ \ }\rangle =\mid (N=0,T=0);K=0,L=0,M=0;T_{0}=0\text{ }\rangle
\label{GS}
\end{equation}
which is the vacuum state for the $Sp(12,R)$ group.

\subsection{\protect\bigskip The symplectic extension of IVBM}

The basis states associated with the even irreducible representation of the $%
Sp(12,R)$ can be constructed by the application of powers of raising
generators $F_{M}^{L}(\alpha ,\beta )$ of the same group. Each raising
operator will increase the number of bosons $N$ by two. As a result we get a
realization of the reduction scheme \cite{sp4nR}:

\begin{equation}
Sp(12,R)\text{ \ }\underrightarrow{{\small N}}\text{\ }U(6)\text{\ \ }%
\underrightarrow{{\small T}^{2}}\text{\ \ \ }SU(2)\times SU(3)\text{\ \ }%
\underrightarrow{{\small T}_{0}}\text{\ \ \ \ }SU(3)
\end{equation}%
The $Sp(12,R)$ classification scheme for the $SU(3)$ boson representations
for even value of the number of bosons $N$ \ is shown on Table 1. Each row
(fixed $N$) of the table corresponds to a given irreducible representation
of the $U(6)$. Then the possible values for the pseudospin are $T=\frac{N}{2}%
,\frac{N}{2}-1,...$ $0$ and are given in the column next to the respective
value of $N$. Thus when $N$ and $T$ are fixed, $2T+1$ equivalent
representations of the group $SU(3)$ arise. Each of them is labelled by the
eigenvalues of the operator $T_{0}:-T,-T+1,...,T,$ defining the columns of
Table 1. The same $SU(3)$ representations $(\lambda ,\mu )$ arise for the
positive and negative eigenvalues of $T_{0}$.

Hence, in the framework of the discussed boson representation of the $%
Sp(12,R)$ algebra all possible irreducible representations of the group $%
SU(3)$ are determined uniquely through all possible sets of the eigenvalues
of the Hermitian operators $N$ and $T^{2}.$ The equivalent use of the $%
(\lambda ,\mu )$ labels facilitates the final reduction to the $SO(3)$
representations, which define the angular momentum $L$ and its projection $%
M. $ The multiplicity index $K$ appearing in this reduction is related to
the projection of $L$ in the body fixed frame and is used with the parity ($%
\pi $)\ to label the different bands ($K^{\pi }$) in the energy spectra of
the nuclei. We define the parity of the states as $\pi =(-1)^{T}$. This
allows us to describe both positive and negative bands.

\subsection{The energy spectrum}

The Hamiltonian, corresponding to the considered, rotational limit of IVBM,\
is expressed in terms of the first and second order invariant operators of
the different subgroups in the chain (\ref{chain}): 
\begin{equation}
H=aN+\alpha _{6}K_{6}+\alpha _{3}K_{3}+\alpha _{1}K_{1}+\beta _{3}\pi _{3},
\label{Hl}
\end{equation}
where $K_{n}$ are the quadratic invariant operators of the $U(n)$ - groups, $%
\pi _{3}$ is \ of the $SO(3)$ second order Casimir operator. As a result of
the connections (\ref{NTcon}) the Casimir operator $K_{3}$ with eigenvalue $%
(\lambda ^{2}+\mu ^{2}+\lambda \mu +3\lambda +3\mu ),$ is expressed in terms
of \ the operators $N$ and $T$: 
\[
K_{3}=2Q^{2}+\frac{3}{4}L^{2}=\frac{1}{2}N^{2}+N+T^{2} 
\]

Making use of the above relation, Hamiltonian (\ref{Hl}) takes the form 
\begin{equation}
H=aN+bN^{2}+\alpha _{3}T^{2}+\beta _{3}\pi _{3}+\alpha _{1}T_{0}^{2},
\end{equation}
and is obviously diagonal in the basis (\ref{bast}) labelled by the quantum
numbers of the subgroups of group-subgroup chain (\ref{chain}). Its
eigenvalues are the energies of the basis states of the boson
representations of $Sp(12,R)$: 
\begin{equation}
E((N,T);KLM;T_{0})=aN+bN^{2}+\alpha _{3}T(T+1)+\beta _{3}L(L+1)+\alpha
_{1}T_{0}^{2}  \label{Erot}
\end{equation}

The \ energy of the \ ground state (\ref{GS}) of the \ system is obviously \ 
$0$.

\section{Application of IVBM for the description of the ground state and
octupole bands energies}

In this paper we modify the earlier application of the IVBM \cite{oepb} for
the description of the first excited even and odd parity bands in order to
reach much higher angular momentum states in both band types. We will apply
the model to even- even deformed nuclei, which exhibit a low-lying negative
parity band next to the ground band traditionally considered to be an
octupole band \cite{revexp}. In order to do this we first have to identify
these experimentally observed bands with the sequences of basis states for
the even representation of $Sp(12,R)$ given in Table 1. We choose the $SU(3)$
multiplet $(0,\mu )$ for a description of the ground band, whereas for the
octupole band the $SU(3)$ multiplet $(2,\mu -1)$ is used. In terms of $(N,T)$
this choice corresponds to $(N=2\mu ,T=0)$ for the positive $(K^{\pi
}=0^{+}) $ and $(N=2\mu +2,T=1)$ for the negative $(K^{\pi }=0^{-})$ parity
band, respectively.

\subsection{ Yrast bands}

In this way, in the framework of the symplectic extension of boson
representations of number preserving $U(6)$ symmetry we are able to consider
\ all even eigenvalues of the number of vector bosons $N$\ with the
corresponding set of pseudospins $T.$

This approach is based on the fact that the energies (\ref{Erot}) increase
with increasing of $N$. We define the energies of each state with \ given $\
L$ \ as yrast energy with respect to $N$ in the two considered$\ $bands$.$
Hence their minimum values \ are obtained at $N=2L$ \ for the ground band,
and $N=2L+2$\ for the octupole band, respectively. So for the description of
the ground band our choice corresponds to the sequence of states \ with
different numbers of bosons, $N=0,4,8,...$ . and pseudospin $T=0$ in the
column labelled $T_{0}=0$ of Table 1. Respectively for a description of the
negative parity band, we choose the set of states \ with quantum numbers $%
N=8,12,...$ and $T=1$ from the same column $T_{0}=0$ . \ Since these quantum
numbers uniquely define the \textbf{\ }$SU(3)$ multiplets, which reduce to
the\ corresponding values of the angular momenta\ $L$, the ground band
belongs to the $SU(3)$ multiplet $(0,\frac{N}{2})$\ and the octupole band to 
$(2,\frac{N}{2}-1)$. In the so defined $SU(3)$ representations for \ each \ $%
N$ \ the maximal values of $L$ appear \ for the first time\ (see Table 1.).\ 

According to the correspondence identified above between the basis states $%
T_{0}=0$ and the experimental data on the ground and octupole bands, the
last term in the energy formula (\ref{Erot}) vanishes. The phenomenological
model parameters $a,b,$ $\alpha _{3},$ and $\beta _{3}$ are evaluated by a
fit to the experimental data. Their values obtained for some even-even
deformed nuclei belonging to light actinides and rare earth region are given
in Table 2. The second column gives the numbers of the experimental states
used in the fitting procedure.

The comparison between the experimental spectra\ and our calculations using
the values of the model parameters given in \ Table 2. for the ground and
octupole bands of the nuclei $Ra^{224},Th^{226}$, $Sm^{152}$ and $Yb^{168}$
is illustrated in Figure 1. All experimental data \ are taken from \ \cite%
{experdata}.

The agreement between the theoretical \ values obtained \ with only \ four
model parameters and the \ experimental data \ for all the \ nuclei under \
consideration is \ rather good.

Applying the yrast conditions relating \ $N$ \ and $\ L$\ \ the energies (%
\ref{Erot}) \ for two considered bands can be rewritten \ as:

\begin{equation}
E(L)=\beta L(L+1)+(\gamma +\eta )L+\xi .  \label{Evr}
\end{equation}
\bigskip The new free parameters $\beta ,\gamma ,\eta ,$ and $\xi $ are
related to the \ previous ones as \ follows:

\begin{equation}
\beta =4b+\beta _{3},\text{ \ }\gamma =2a-4b,\text{ \ \ }\eta =8b,\text{ \ \ 
}\xi =2a+4b+2\alpha _{3}.  \label{Lpar}
\end{equation}

The values of $\beta $ and $\gamma $ can be determined only from a fit to
the positive band energies,\ while $\eta $ and $\xi $ are \ estimated \ from
the negative ones, respectively. The values of the parameters (\ref{Lpar})\
determine the behavior of the energies of \ the two bands and their
position\ with respect to each other. In some cases ($%
^{232}Th,^{234}U,^{236}U,^{238}U$) the two bands are almost parallel. The
shift between them depends on the parameter $\xi .$ When they are very close
they interact through the $L$ -dependent interaction with a strength $\gamma
+\eta .$

As a result of our theoretical assumptions we obtained a simple formula for
the energy levels. From (\ref{Evr}) we can see that eigenstates of the first
positive and negative bands consists of rotational \ $L(L+1)$\ and
vibrational $L$\ modes. The rotational \ interaction is with equal strength
\ $\beta $ in both of the bands. The obtained values of the parameter $\eta $
are always negative, which means that the negative parity band is less
vibrational than the positive one.

\subsection{The staggering}

In the collective rotational spectra of deformed even-even nuclei in this
mass region some fine structure effects as back-banding and staggering
behavior are observed . Odd-even staggering patterns between ground and
octupole bands have been investigated recently \cite{NS2}. In order to test
further our model we applied on the energies the staggering function defined
as \cite{stag}:

\begin{equation}
Stg(L)=6\Delta E(L)-4\Delta E(L-1)-4\Delta E(L+1)+\Delta E(L+2)+\Delta
E(L-2),
\end{equation}%
where $\Delta E(L)=E(L)-E(L-1).$ This function is a finite difference of
fourth order in respect to $\Delta E(L)$ or of fifth order in respect to
energy $E(L)$ and is characteristic for the deviation of the rotational
behavior from that of the rigid rotor. The calculated and experimental
staggering patterns are illustrated in Figure 2. One can see a good
agreement \ with experiment, as well as the reproduction of the
\textquotedblright beat\textquotedblright\ patterns of \ the staggering
behavior. \ \ They occur in the \ region where the \ interaction \ between
the\ two considered \ bands is most strong \ or they cross. The correct
reproduction of the experimental staggering patterns is \ due \ to the
interaction term $\eta L$ \ in (\ref{Evr}) between the positive and negative
parity bands, which is a result of the introduced notion of yrast energies
in the framework of the symplectic extension of the IVBM.

\section{\protect\bigskip \protect\bigskip Conclusions}

We have applied the Interacting Vector Boson Model for the description of
the ground and octupole bands in some even-even rare earth and actinide
nuclei up to very high spins. In spite of the simplicity of the model
without introducing additional degrees of freedom we are \ able to describe
\ both positive and \ negative parity bands. This \ is \ due to the specific
definition of the states parity depending \ on the pseudospin quantum number 
$T$.

The successful reproduction of the experimental energies and of their
odd-even staggering was achieved as a result of their consideration as yrast
energies in respect to the number of phonon excitation $N$ \ that build the
collective states. The introduction of this notion was possible, as we
extended the IVBM to its symplectic dynamical symmetry \ $Sp(12,R)$, which
allows the change of the number of bosons that are the building blocks of
the model Hamiltonian. Nevertheless \ the Hamiltonian remains with only few
phenomenological parameters and is still exactly solvable. Through the
algebraic properties of the dynamical symmetry chain relations between $%
SU(3) $ and $U(2)$ quantum numbers are \ established. Combining these
relations with the notion of yrast \ energies the physical meaning of each
term of the Hamiltonian is clarified. In the rotational limit of the model
in addition to the rotational character of the considered bands an purely
vibrational mode is appearing, which introduces also some interaction
between them. This is the reason for the reproduction also of the fine
effect of the structure of these bands. The obtained physically meaningful
results are also simple and easy for use and they permit the application of
the model to larger class of nuclei than the purely rotational ones.

The symplectic extension of the Interacting \ Vector Boson Model permits a
richer classification of the states than its unitary \ version and gives the
possibility for a further consideration of other collective bands. In
general the \ model proves \ appropriate for the \ description of diverse
nuclear structure problems.

\section{\protect\bigskip Acknowledgements}

The authors are grateful for fruitful \ discussions and help \ on \ the \
subject of this paper \ to professors J. P. Draayer, J. Cseh and D.
Bonatsos. This work \ was partially \ supported by \ Bulgarian Science
Committee under \ contract number $\Phi -905$.

\[
\begin{tabular}{lllllll}
\textbf{Table 1.} &  &  &  &  &  &  \\ \hline\hline
\multicolumn{1}{||l}{$N$ (T)$\backslash $ $T_{_{0}}$} & \multicolumn{1}{||l}{%
$...$} & \multicolumn{1}{||l}{$\pm 4$} & \multicolumn{1}{||l}{$\pm 3$} & 
\multicolumn{1}{||l}{$\pm 2$} & \multicolumn{1}{||l|}{$\ \pm 1$} & 
\multicolumn{1}{||l||}{$\ \ 0$} \\ \hline\hline
\multicolumn{1}{||l}{$0%
\begin{tabular}{l}
$0$%
\end{tabular}%
\ \ $} & \multicolumn{1}{||l}{} &  &  &  &  & \multicolumn{1}{|l|}{%
\begin{tabular}{l}
$(0,0)$%
\end{tabular}%
} \\ \cline{1-1}\cline{6-7}
\multicolumn{1}{||l}{$2%
\begin{tabular}{l}
$1$ \\ 
$0$%
\end{tabular}%
\ \ $} & \multicolumn{1}{||l}{} &  &  &  & \multicolumn{1}{|l|}{%
\begin{tabular}{l}
$(2,0)$%
\end{tabular}%
} & \multicolumn{1}{|l|}{%
\begin{tabular}{l}
$(2,0)$ \\ 
$(0,1)$%
\end{tabular}%
} \\ \cline{1-1}\cline{5-7}
\multicolumn{1}{||l}{$4%
\begin{tabular}{l}
$2$ \\ 
$1$ \\ 
$0$%
\end{tabular}%
\ \ $} & \multicolumn{1}{||l}{} &  &  & \multicolumn{1}{|l}{%
\begin{tabular}{l}
$(4,0)$%
\end{tabular}%
} & \multicolumn{1}{|l|}{%
\begin{tabular}{l}
$(4,0)$ \\ 
$(2,1)$%
\end{tabular}%
} & \multicolumn{1}{|l|}{%
\begin{tabular}{l}
$(4,0)$ \\ 
$(2,1)$ \\ 
$(0,2)$%
\end{tabular}%
} \\ \cline{1-1}\cline{4-6}\cline{6-7}
\multicolumn{1}{||l}{$6%
\begin{tabular}{l}
$3$ \\ 
$2$ \\ 
$1$ \\ 
$0$%
\end{tabular}%
\ \ $} & \multicolumn{1}{||l}{} &  & \multicolumn{1}{|l}{$%
\begin{tabular}{l}
$(6,0)$%
\end{tabular}%
\ \ $} & \multicolumn{1}{|l}{%
\begin{tabular}{l}
$(6,0)$ \\ 
$(4,1)$%
\end{tabular}%
} & \multicolumn{1}{|l|}{%
\begin{tabular}{l}
$(6,0)$ \\ 
$(4,1)$ \\ 
$(2,2)$%
\end{tabular}%
} & \multicolumn{1}{|l|}{%
\begin{tabular}{l}
$(6,0)$ \\ 
$(4,1)$ \\ 
$(2,2)$ \\ 
$(0,3)$%
\end{tabular}%
} \\ \cline{1-1}\cline{3-7}
\multicolumn{1}{||l}{$8%
\begin{tabular}{l}
$4$ \\ 
$3$ \\ 
$2$ \\ 
$1$ \\ 
$0$%
\end{tabular}%
\ \ $} & \multicolumn{1}{||l}{} & \multicolumn{1}{|l}{%
\begin{tabular}{l}
$(8,0)$%
\end{tabular}%
} & \multicolumn{1}{|l}{%
\begin{tabular}{l}
$(8,0)$ \\ 
$(6,1)$%
\end{tabular}%
} & \multicolumn{1}{|l}{$%
\begin{tabular}{l}
$(8,0)$ \\ 
$(6,1)$ \\ 
$(4,2)$%
\end{tabular}%
\ \ $} & \multicolumn{1}{|l|}{%
\begin{tabular}{l}
$(8,0)$ \\ 
$(6,1)$ \\ 
$(4,2)$ \\ 
$(2,3)$%
\end{tabular}%
} & \multicolumn{1}{|l|}{%
\begin{tabular}{l}
$(8,0)$ \\ 
$(6,1)$ \\ 
$(4,2)$ \\ 
$(2,3)$ \\ 
$(0,4)$%
\end{tabular}%
} \\ \cline{1-1}\cline{3-7}
\multicolumn{1}{||l}{$...$} & \multicolumn{1}{||l}{$...$} & 
\multicolumn{1}{|l}{$...$} & \multicolumn{1}{|l}{$...$} & 
\multicolumn{1}{|l}{...} & \multicolumn{1}{|l|}{...} & \multicolumn{1}{|l|}{
...} \\ \cline{3-3}
\end{tabular}%
\ \ 
\]

\[
\begin{tabular}{llllll}
\textbf{Table 2.} &  &  &  &  &  \\ \hline\hline
\multicolumn{1}{||l}{Nucleus} & \multicolumn{1}{||l}{$n_{s}$} & 
\multicolumn{1}{||l|}{$a$} & \multicolumn{1}{||l|}{$b$} & 
\multicolumn{1}{||l|}{$\alpha _{3}$} & \multicolumn{1}{||l||}{$\beta _{3}$}
\\ \hline\hline
\multicolumn{1}{||l}{$Ra^{224}$} & \multicolumn{1}{||l}{\small 13} & 
\multicolumn{1}{|l}{\small 0.0119} & \multicolumn{1}{|l}{\small -0.0022} & 
\multicolumn{1}{|l}{\small 0.0789} & \multicolumn{1}{|l||}{\small 0.0155} \\ 
\hline
\multicolumn{1}{||l}{$Ra^{226}$} & \multicolumn{1}{||l}{\small 18} & 
\multicolumn{1}{|l}{\small 0.0269} & \multicolumn{1}{|l}{\small -0.0005} & 
\multicolumn{1}{|l}{\small 0.0226} & \multicolumn{1}{|l||}{\small 0.0060} \\ 
\hline
\multicolumn{1}{||l}{$Th^{222}$} & \multicolumn{1}{||l}{\small 26} & 
\multicolumn{1}{|l}{\small 0.0558} & \multicolumn{1}{|l}{\small 0.0000} & 
\multicolumn{1}{|l}{\small -0.0557} & \multicolumn{1}{|l||}{\small 0.0030}
\\ \hline
\multicolumn{1}{||l}{$Th^{224}$} & \multicolumn{1}{||l}{\small 18} & 
\multicolumn{1}{|l}{\small 0.0242} & \multicolumn{1}{|l}{\small -0.0011} & 
\multicolumn{1}{|l}{\small 0.0362} & \multicolumn{1}{|l||}{\small 0.0100} \\ 
\hline
\multicolumn{1}{||l}{$Th^{226}$} & \multicolumn{1}{||l}{\small 20} & 
\multicolumn{1}{|l}{\small 0.0194} & \multicolumn{1}{|l}{\small -0.0009} & 
\multicolumn{1}{|l}{\small 0.0522} & \multicolumn{1}{|l||}{\small 0.0094} \\ 
\hline
\multicolumn{1}{||l}{$Th^{228}$} & \multicolumn{1}{||l}{\small 18} & 
\multicolumn{1}{|l}{\small 0.0092} & \multicolumn{1}{|l}{\small -0.0020} & 
\multicolumn{1}{|l}{\small 0.1470} & \multicolumn{1}{|l||}{\small 0.0138} \\ 
\hline
\multicolumn{1}{||l}{$Th^{232}$} & \multicolumn{1}{||l}{\small 29} & 
\multicolumn{1}{|l}{\small 0.0155} & \multicolumn{1}{|l}{\small -0.0021} & 
\multicolumn{1}{|l}{\small 0.3244} & \multicolumn{1}{|l||}{\small 0.0128} \\ 
\hline
\multicolumn{1}{||l}{$U^{234}$} & \multicolumn{1}{||l}{\small 19} & 
\multicolumn{1}{|l}{\small 0.0124} & \multicolumn{1}{|l}{\small -0.0010} & 
\multicolumn{1}{|l}{\small 0.3608} & \multicolumn{1}{|l||}{\small 0.0085} \\ 
\hline
\multicolumn{1}{||l}{$U^{236}$} & \multicolumn{1}{||l}{\small 25} & 
\multicolumn{1}{|l}{\small 0.0154} & \multicolumn{1}{|l}{\small -0.0010} & 
\multicolumn{1}{|l}{\small 0.2846} & \multicolumn{1}{|l||}{\small 0.0086} \\ 
\hline
\multicolumn{1}{||l}{$U^{238}$} & \multicolumn{1}{||l}{\small 27} & 
\multicolumn{1}{|l}{\small 0.0142} & \multicolumn{1}{|l}{\small -0.0016} & 
\multicolumn{1}{|l}{\small 0.2851} & \multicolumn{1}{|l||}{\small 0.0110} \\ 
\hline
\multicolumn{1}{||l}{$Yb^{168}$} & \multicolumn{1}{||l}{\small 41} & 
\multicolumn{1}{|l}{\small 0.0235} & \multicolumn{1}{|l}{\small -0.0056} & 
\multicolumn{1}{|l}{\small 0.6512} & \multicolumn{1}{|l||}{\small 0.0295} \\ 
\hline
\multicolumn{1}{||l}{$Sm^{152}$} & \multicolumn{1}{||l}{\small 15} & 
\multicolumn{1}{|l}{\small 0.0194} & \multicolumn{1}{|l}{\small -0.0045} & 
\multicolumn{1}{|l}{\small 0.4290} & \multicolumn{1}{|l||}{\small 0.0274} \\ 
\hline\hline
\end{tabular}%
\ \ 
\]


\bigskip

\begin{thebibliography}{99}
\bibitem{revexp} P. A. Butler and W. Nazarewicz, Rev. of Mod. Phys. \emph{%
\textbf{\ }}\textbf{68}\emph{\ }(1996) 349.

\bibitem{IBM} F. Iachello, A. Arima, \emph{''The Interacting Boson Model'',}
Cambridge University Press, Cambridge, 1987.

\bibitem{Iach2} J. Engel and F. Iachello, Phys. Rev. Lett. \textbf{54}
(1985), 1126; J. Engel and F. Iachello, Nucl. Phys. \textbf{A472} (1987) 61;
V. \ Zamfir \ and \ D. Kusnezov Phys. Rev. \textbf{C 63 }(2001) 054306.

\bibitem{alonso} Alonso et al., Nucl. Phys. \textbf{A586} (1995) 100.

\bibitem{raduta} A. A. Raduta and D. Ionescu, Phys. Rev. \textbf{C 67}
(2003) 044312.

\bibitem{NS1} N. Minkov, S. Drenska, P. Raychev, R. Roussev, and D.
Bonatsos, Phys. Rev. \textbf{C 63 }(2001) 044305.

\bibitem{NS2} N. Minkov, S. Drenska, Prog. Theor. Phys. Suppl. \textbf{146 }%
(2002) 597.

\bibitem{IVBMb} A. Georgieva, P. Raychev, and R. Roussev,\emph{\ {J. Phys.
G: Nucl. Phys.} \ }\textbf{8}\emph{\ }(1982) 1377.

\bibitem{RuRa} P. Raychev, and R. Roussev, \emph{J. Phys. G: Nucl. Phys. }%
\textbf{\ 7}\emph{\ }(1981) 1227.

\bibitem{NS3} N. Minkov, S. Drenska, P. Raychev, R. Roussev, and Dennis
Bonatsos, Phys. Rev. \textbf{C 55} (1997) 2345.

\bibitem{IVBMrl} A. Georgieva, P. Raychev, and R. Roussev, \emph{J. Phys. G:
Nucl. Phys. }\textbf{\ 9}\emph{\ }(1983) 521.

\bibitem{oepb} A. Georgieva, P. Raychev, and R. Roussev,\emph{\ Interacting
Vector Boson Model and Negative Parity Bands in Actinides, Bulg. J. Phys.}%
\textbf{\ 12}\emph{\ }(1985) 2 147.

\bibitem{ggg} H.Ganev, V. Garistov, A. Georgieva, Proceedings of the XXII
International Workshop on Nuclear Theory, Rila Mountains, Bulgaria, June
15-21 2003.

\bibitem{sp4nR} A. Georgieva, M. Ivanov, P. Raychev, and R.\ Roussev, 
\textit{\ Int.J. Theor. Phys. }\textbf{28 \ }(1989) 769.

\bibitem{experdata} Mitsuo Sacai Atomic Data and Nuclear Data Tables 31,
399-432 (1984); \textbf{Level Retrieval Parameters \ }%
http://iaeand.iaea.or.at/nudat/levform.html

\bibitem{stag} D. Bonatsos, C. Daskaloyannis, S. Drenska, N. Kaoussos, N.
Minkov, P. Raychev, and R. Roussev, Phys. Rev. \textbf{C62} (2000) 024301.

\newpage
\end{thebibliography}
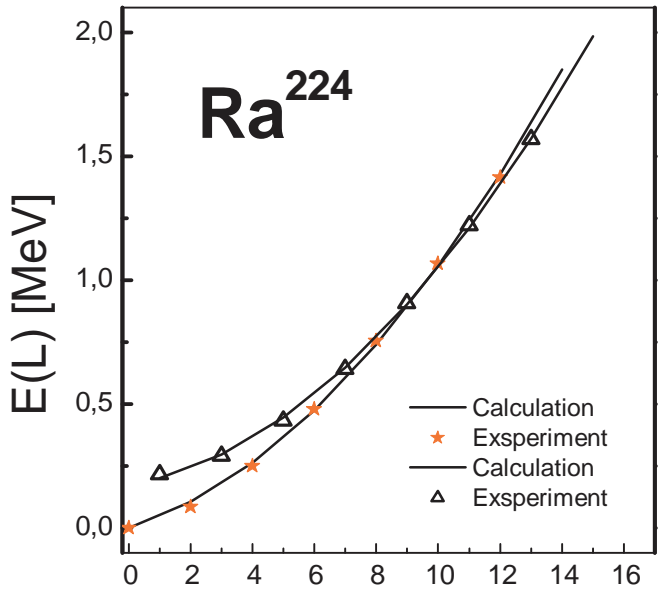
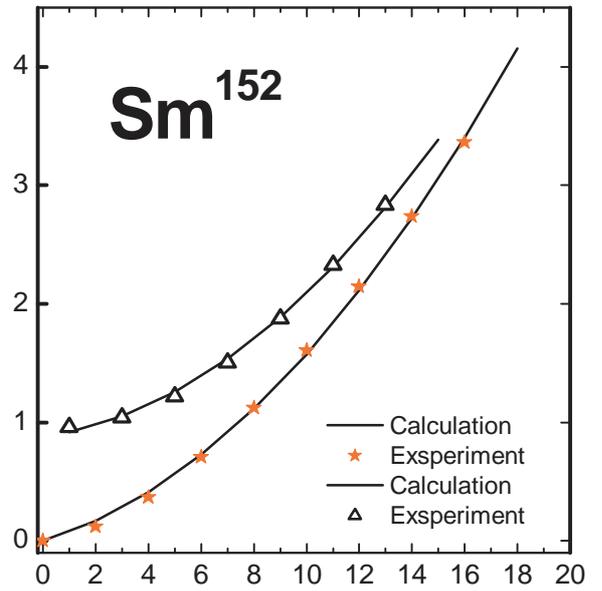
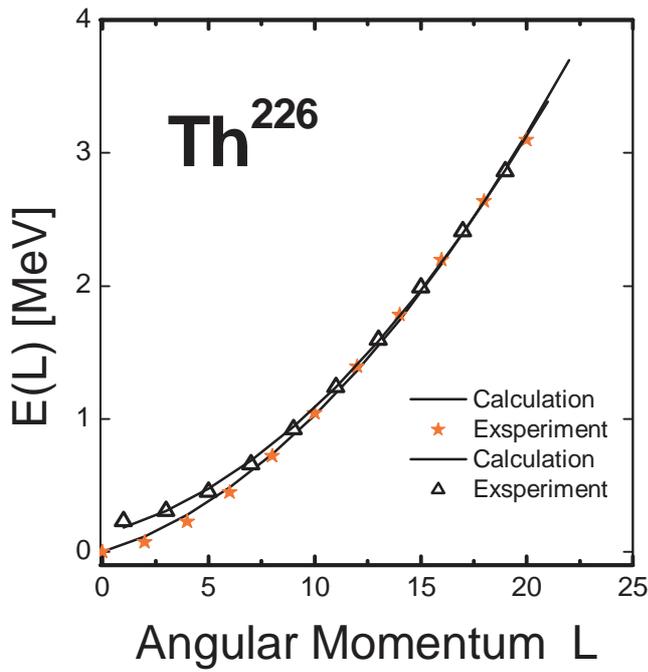
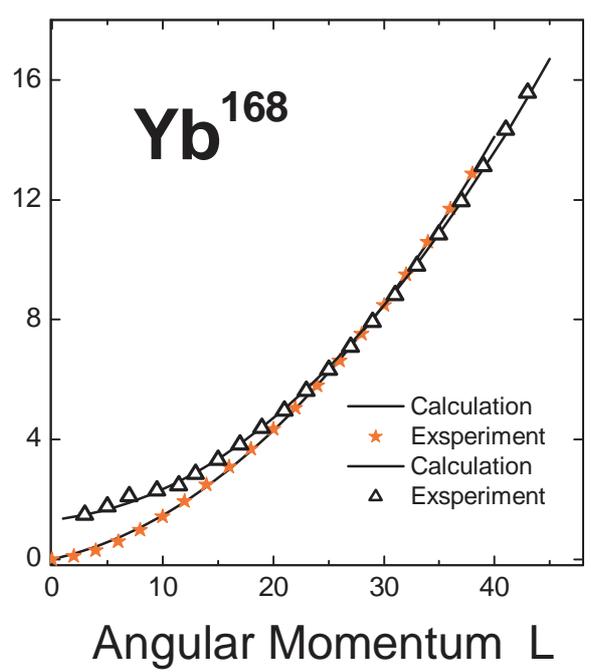
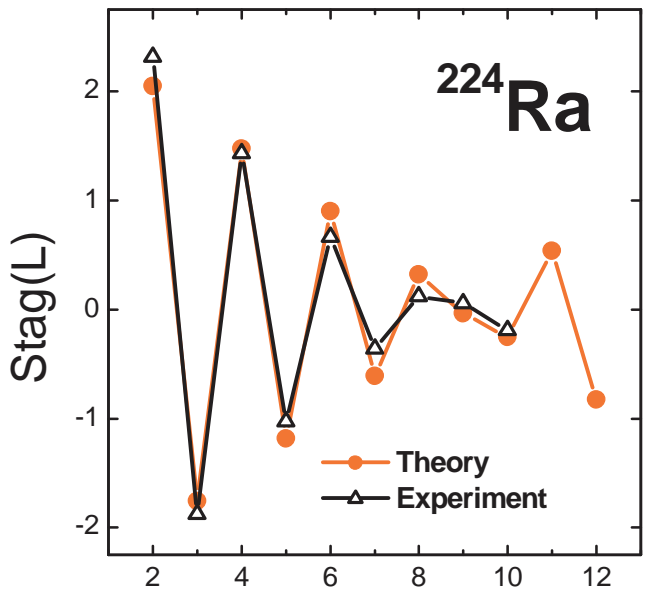
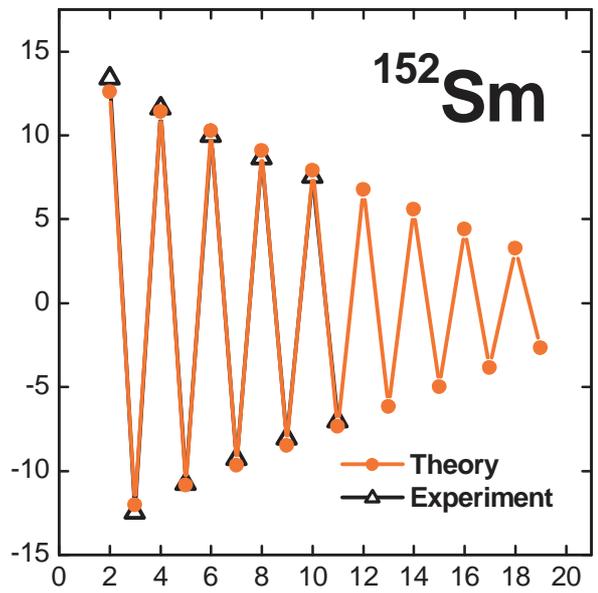
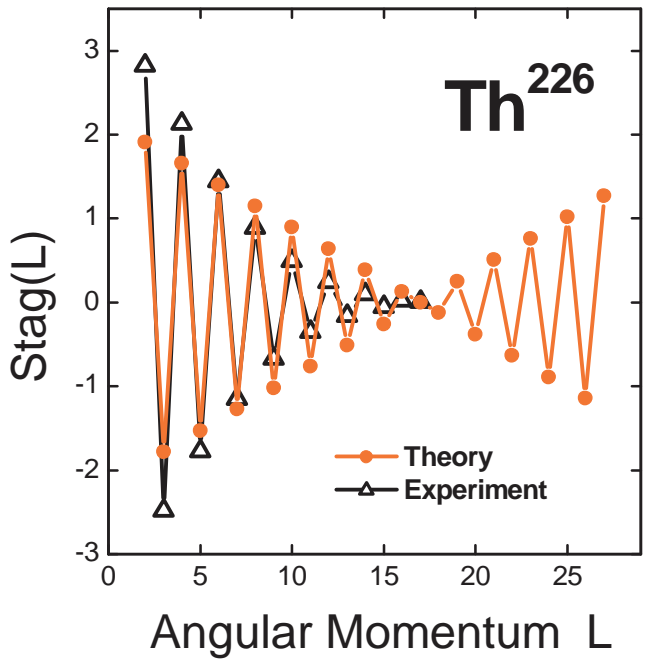
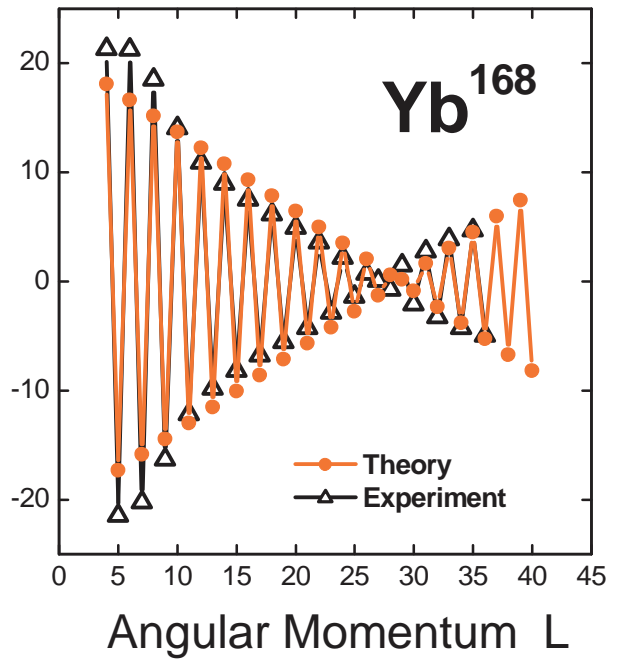
\end{document}